# Testing CCC+TL Cosmology with Observed BAO Features

Rajendra P. Gupta[1]

*Department of Physics, University of Ottawa, Ottawa, Canada*

**ABSTRACT**

The primary purpose of this paper is to see how well a recently proposed new model fits (a) the position of the baryon acoustic oscillations (BAO) features observed in the large-scale distribution of galaxies and (b) the angular size measured for the sound horizon due to BAO imprinted in the cosmic microwave background (CMB) anisotropy. The new model is a hybrid model that combines the tired light (TL) theory with a variant of the ΛCDM model in which the cosmological constant is replaced with a covarying coupling constants' (CCC) parameter $\alpha$. This model, dubbed the CCC+TL model, can fit the supernovae type 1a Pantheon+ data as accurately as the ΛCDM model, and also fit the angular size of cosmic dawn galaxies observed by the James Webb Space Telescope, which is in tension with the ΛCDM model. The results we obtained are 151.0 (±5.1) Mpc for the absolute BAO scale at the current epoch, and the angular size of the sound horizon $\theta_{sh} = 0.60°$ matching Planck's observations at the surface of the last scattering when the baryon density is set to 100% of the matter density and $|\alpha|$ is increased by 5.6%. It remains to be seen if the new model is consistent with the CMB power spectrum, the big-bang nucleosynthesis of light elements, and other critical observations.

**Keywords**: Cosmology, Galaxies, (cosmology:) early Universe < Cosmology, galaxies: high-redshift < Galaxies, baryon acoustic oscillations, CMB, sound horizon

## I. INTRODUCTION

One of the most important tests of a cosmological model is to reproduce in the galaxy distribution the signatures of the baryon acoustic oscillations (BAO) resulting from the sound waves in the baryon-photon fluid at the time when photons and baryon decoupled, and radiation traveled freely in space. This radiation is observed as the cosmic microwave background (CMB). The BAO features are detected as tiny temperature fluctuations (anisotropies) in the highly isotropic CMB observations. These oscillations are believed to develop into large-scale structures as higher-density regions of perturbations become the nucleation points where galaxies form. Thus, the two features are expected to be correlated, and BAO could serve as a fundamental standard ruler to test cosmological models (Peebles and Yu 1970; Bond and Efstathiou 1984; Eisenstein and Hu 1998; Meiksin et al. 1999). The observation of BAO features at different redshifts endorses the propagation of primordial gravitational instability (Eisenstein et al. 2005; Cole et al. 2005; Fronenberg et al. 2023). Xu et al. (2023) have recently established evidence for baryon acoustic oscillations from galaxy–ellipticity correlations. Tully et al. (2023) have presented remarkably strong evidence for the existence of *individual* BAO signal at $z = 0.068$. Shao et al. (2023) raised the possibility of using the angular scale of cosmic inhomogeneities as a new, model-independent way to constrain cosmological parameters.

Sutherland (2012) succinctly stated: '… the observed BAO features support the standard cosmology in several independent ways: the existence of features supports the basic gravitational instability paradigm for structure formation; the relative weakness of BAO feature supports the ~1:5 ratio of baryon to dark matter, since baryon dominated universe would have a BAO feature much stronger than observed; and the observed length-scale of the feature in redshift space is consistence with the concordance ΛCDM model derived from the CMB and other observations, with $\Omega_m \approx 0.27$ and $H_0 \approx 70 \text{ km s}^{-1} \text{ Mpc}^{-1}$ (Komatsu et al. 2011)'.

Weinberg et al. (2012) have shown that BAO features in the matter power spectrum (galaxies) when combined with the tracer power spectrum (CMB), can effectively constrain the cosmological parameter and test a model. For our purpose, we are interested in the BAO measurements of angular separations $\theta_{BAO}$ of pairs of galaxies at different redshift values. Carvalho et al. (2016) used 409,337 luminous red galaxies in the redshift range $z = [0.440, 0.555]$ to estimate $\theta_{BAO}(z)$ at six redshift shells. Their work was extended to include observations that provided $\theta_{BAO}$ at $z = 0.11$ within $z = [0.105, 0.115]$ (Carvalho et al. 2021), and at up to $z = 0.65$ (Lemos et al. 2023). We will use their data to explore if the recently proposed hybrid model (Gupta 2023) is consistent with features of the baryon acoustic oscillations (BAO) observed in the cosmic microwave background (CMB) and matter power spectra. This model comprises a modified ΛCDM model permitting the covariation of coupling constants (CCC) and includes the tired light (TL) phenomenon to partially account for the observed redshift. This two-parameter hybrid model, dubbed CCC+TL, was able to account for the bewildering observation by the James Webb Space Telescope showing unexpected morphology of galaxies existent at cosmic dawn. The model parameters $H_0$ and $\alpha$, the latter determining the strength of the coupling constants' variation and replaces Λ of the standard model, are determined by fitting the supernovae type 1a (Pantheon+) data (Scolnic et al. 2022, Brout et al. 2022).

---

[1] Email: rgupta4@uottawa.ca

The galaxies observed in the early universe, some less than 500 million years after the Big Bang, appear to have shapes, structures, and masses similar to those in existence for billions of years (e.g., Naidu et al. 2022a, 2022b; Labbe et al. 2023; Curtis-Lake et al. 2023; Hainline et al. 2023; Robertson et al. 2023) but with angular sizes an order of magnitude smaller than expected for such galaxies (e.g., Adams et al. 2022, Atek et al. 2022, Chen et al. 2022, Donnan et al. 2022, Finkelstein et al. 2022, Naidu et al. 2022a, 2022b, Ono et al. 2022, Tacchella et al. 2022, 2023; Wu et al. 2022, Yang et al. 2022, Austin et al. 2023, Baggen et al. 2023). Attempts have been made to resolve the problem by modifying the star and galaxy formation models (e.g., Haslbauer et al. 2022; Inayoshi et al. 2022; Kannan et al. 2022; Keller et al. 2022; Regan 2022; Yajima et al. 2022; Atek et al. 2023, Mason et al. 2023; Mirocha & Furlanetto 2023; Whitler et al. 2023a, 2023b; McCaffrey et al. 2023), such as by compressing time for the formation of population III stars and galaxies more and more by considering the presence of primordial massive black hole seeds, and super-Eddington accretion rates in the early Universe (Ellis 2022, Bastian et al. 2023, Brummel-Smith 2023, Chantavat et al. 2023, Dolgov 2023, Larson et al. 2023, Maiolino et al. 2023). Other researchers (Dekel et al. 2023; Boyett et al. 2023; Looser et al. 2023; Long et al. 2023; Bunker et al. 2023; Haro et al. 2023; Eilers et al. 2023;) are concerned if they provide satisfactory answers. Some even suggest looking for new physics (Schneider et al. 2023; Chen et al. 2023; Mauerhofer and Dayal 2023). In the words of Garaldi et al.(2023): 'Cosmological simulations serve as invaluable tools for understanding the Universe. However, the technical complexity and substantial computational resources required to generate such simulations often limit their accessibility within the broader research community. Notable exceptions exist, but most are not suited for simultaneously studying the physics of galaxy formation and cosmic reionization during the first billion years of cosmic history.' According to Xiao et al. (2023): 'Massive optically dark galaxies unveiled by JWST challenge galaxy formation models.' (See also Katz et al. 2023; Ormerod et al. 2023; Greene et al. 2023).

The CCC+TL model predicts the age of the universe as 26.7 Gyr against the generally accepted value of 13.8 Gyr. This is of deep concern and needs the model validation against multiple observations, including BAO, CMB, BBN (big-bang nucleosynthesis), and globular cluster ages. Our focus here is on BAO. This paper is organized to include the Theoretical Background in Section II, Results in Section III, Discussion in Section IV, and Conclusions in Section V.

## II. THEORETICAL BACKGROUND

*CCC Model:* The modified FLRW metric, incorporating the covarying coupling constant (CCC) concept, is (Gupta 2023)[2]

$$ds^2 = c_0^2 dt^2 f(t)^2 - a(t)^2 f(t)^2 \left( \frac{dr^2}{1-kr^2} + r^2(d\theta^2 + \sin^2\theta \, d\phi^2) \right), \quad (1)$$

the Friedmann equations are

$$\left(\frac{\dot{a}}{a} + \alpha\right)^2 = \frac{8\pi G_0}{3c_0^2}\varepsilon - \frac{kc_0^2}{a^2}, \text{ and} \quad (2)$$

$$\frac{\ddot{a}}{a} = -\frac{4\pi G_0}{3c_0^2}(\varepsilon + 3p) - \alpha\left(\frac{\dot{a}}{a}\right), \quad (3)$$

and the continuity equation is

---

[2] For the sake of clarity we have repeated some of the material from Gupta (2023).

$$\dot{\varepsilon} + 3\frac{\dot{a}}{a}(\varepsilon + p) = -\alpha(\varepsilon + 3p). \tag{4}$$

Here $a$ is the scale factor, $G_0$ is the current value of the gravitational constant, $c_0$ is the current value of the speed of light, $k$ is the curvature constant, $\alpha$ is a constant defining the variation of the constant through a function $f(t) = \exp(\alpha(t-t_0))$ with time measured from the beginning of the universe and $t_0$ the current time, $\varepsilon$ is the energy density of all the components, and $p$ is their pressure. Using the function $f(t)$, $c(t) = c_0 f(t)$ and $G = G_0 f(t)^3$ in the CCC model. Solution of this Eq. (4) for matter ($p = 0$) and radiation ($p = \varepsilon/3$) are, respectively,

$$\varepsilon_m = \varepsilon_{m,0} a^{-3} f^{-1}, \text{ and } \varepsilon_r = \varepsilon_{r,0} a^{-4} f^{-2}. \tag{5}$$

Defining the Hubble expansion parameter as $H = \dot{a}/a$, we may write Eq. (2) for a flat universe ($k = 0$) as

$$(H + \alpha)^2 = \frac{8\pi G_0}{3c_0^2} \varepsilon \Rightarrow \varepsilon_{c,0} \equiv \frac{3c_0^2(H_0+\alpha)^2}{8\pi G_0}. \tag{6}$$

This equation defines the critical density of the universe in the CCC model. Using Eqs. (5) and (6), we may write

$$H = (H_0 + \alpha)\left(\Omega_{m,0} a^{-3} f^{-1} + \Omega_{r,0} a^{-4} f^{-2}\right)^{1/2} - \alpha. \tag{7}$$

In this equation, relative matter density $\Omega_{m,0} = \varepsilon_{m,0}/\varepsilon_{c,0}$ and relative radiation density $\Omega_{r,0} = \varepsilon_{r,0}/\varepsilon_{c,0}$. Since $\Omega_{r,0} \ll \Omega_{m,0}$, and we don't have to worry about the dark energy density in the CCC model, Eq. (7) simplifies to

$$H = (H_0 + \alpha)\left(a^{-3} f^{-1} + \Omega_{r,0} a^{-4} f^{-2}\right)^{1/2} - \alpha. \tag{8}$$

Since the observations are made using redshift $z$, we have to see how the scale factor $a$ relates to $z$ in the CCC model. Along the spatial geodesic ($\theta$ and $\phi$ constant) between the observer and the source at a fixed time $t$ using the modified FLRW metric (Eq. 1)

$$ds = a(t)f(t)dr. \tag{9}$$

Thus, the proper distance for commoving coordinate $r$ [since $a(t_0) \equiv 1 = f(t_0)$]

$$d_p = a(t)f(t) \int_0^r dr = a(t)f(t)r \Rightarrow d_p(t_0) = r. \tag{10}$$

Since light follows the null geodesic, Eq. (1) yields for a light emitted by a source at a time $t_e$ and detected by the observer at a time $t_0$

$$c_0 \int_{t_e}^{t_0} \frac{dt}{a(t)} = \int_0^r dr = r = d_p(t_0). \tag{11}$$

It can now be easily shown (e.g., Ryden 2017) that $a = 1/(1 + z)$, i.e., the same as for the $\Lambda$CDM model.

We now need to transpose $f(t)$ to $f(z)$, as it is the latter that we will require in calculating the proper distance. Following Gupta (2023, Eqs. 24-27), we have

$$f^{-1/2} = x = \left(-\frac{D}{2A} + \left(\left(-\frac{D}{2A}\right)^2 + \left(\frac{C}{3A}\right)^3\right)^{1/2}\right)^{1/3} + \left(-\frac{D}{2A} - \left(\left(-\frac{D}{2A}\right)^2 + \left(\frac{C}{3A}\right)^3\right)^{1/2}\right)^{1/3}, \text{ where}$$

$$A = 1 - \frac{3}{2}\frac{(H_0+\alpha)}{\alpha} = 1 - C, D = -a^{3/2}, \text{ and } x = \exp\left(-\frac{\alpha(t-t_0)}{2}\right). \tag{12}$$

Since the scale factor $a = 1/(1 + z)$, we have $D = -[1/(1 + z)]^{3/2}$. Thus, the above equation provides the function $f(z, H_0, \alpha)$ for the matter-dominated universe. What about its expression in the radiation-dominated universe? Since $z \gg 1$ in such a universe, i.e., $t \ll t_0$, it is easy to see that $f(t) = \exp(\alpha(t - t_0))$ approaches a constant value. We can, therefore, use the same expression for $f(z)$, i.e., Eq. (12) for all values of $z$.

We now need the expression for the proper distance $d_p$. Since $dt = dt \times da/da = da/\dot{a}$, we may write Eq. (10)

$$d_p(t_0) = c_0 \int_{t_e}^{t_0} \frac{dt}{a(t)} = c_0 \int_{a_e}^{1} \frac{da}{a\dot{a}} = c_0 \int_{a_e}^{1} \frac{da}{a^2 H}, \tag{13}$$

and since $a = 1/(1+z), da = -dz/(1+z)^2 = -dza^2$, we get, using Eq. (8)

$$d_p(t_0) = c_0 \int_0^z \frac{dz}{H} = c_0 \int_0^z \frac{dz}{(H_0+\alpha)((1+z)^3 f^{-1}+\Omega_{r,0}(1+z)^4 f^{-2})^{1/2} - \alpha}. \tag{14}$$

We can now follow TL and CCC+TL models developed by Gupta (2023) for our computations in this work.

**Sound horizon distance and angular size**: Denoted as $d_{sh}(t_{ls})$, it represents the distance sound travels at the speed $c_s(t)$ in the photon-baryon fluid from the big-bang until such plasma cools down and disappears due to the formation of atoms, i.e., until the time of last scattering $t_{ls}$ corresponding to the redshift $z_{ls}$. One may write the sound horizon distance (Durrer 2008) using the metric given by Eq. (1) for the CCC+TL model,

$$d_{sh}(t_{ls}) = a(t_{ls}) \int_0^{t_{ls}} \frac{c_s(t)dt}{a(t)}. \tag{15}$$

The speed of sound in terms of the speed of light given by (Durrer 2008),

$$c_s(t) \approx \frac{c_0}{\sqrt{3}} \left(1 + \frac{3\Omega_b}{4\Omega_r}\right)^{-1/2} = \frac{c_0}{\sqrt{3}} \left(1 + \frac{3\Omega_{b,0}}{4\Omega_{r,0}} a(t)f(t)\right)^{-1/2}, \text{ or} \tag{16}$$

$$c_s(z) \approx \frac{c_0}{\sqrt{3}} \left(1 + \frac{3\Omega_{b,0}f(z)}{4\Omega_{r,0}(1+z)}\right)^{-1/2}, \tag{17}$$

where $\Omega_b$ is the baryon density and $\Omega_r$ is the radiation density. Following Eq. (13) and (14), we get

$$d_{sh}(z_{ls}) = \frac{1}{1+z_{ls}} \int_\infty^{z_{ls}} \frac{c_s(z)dz}{(H_c+\alpha)((1+z)^3 f^{-1}+\Omega_{r,0}(1+z)^4 f^{-2})^{1/2} - \alpha}. \tag{18}$$

It is to be noted that the variation of the speed of light is already included in the Friedmann equations and all the expressions derived from it.

Next, we have to determine $z_{ls}$ in the CCC+TL model wherein the constants evolve as $c(t) = c_0 f(t)$, $G = G_0 f(t)^3$, $h = h_0 f(t)^2$, and $k_B = k_{B,0} f(t)^2$, and distance is measured using the speed of light (Gupta 2022a). How does the CMB Planck spectrum evolve with the redshift? We know that the frequency evolves as $\nu = \nu_0(1 + z), d\nu = d\nu_0(1 + z)$, and volume $V$ evolves as,

$$V = V_0 a(z_c)^3 f(z_c)^3 = V_0 (1+z_c)^{-3} f(z_c)^3 = V_0 (1+z)^{-3} f(z_c)^3 (1+z_t)^3, \tag{19}$$

since $1 + z = (1+z_c)(1+z_t)$. The energy density of the CMB photons in the frequency range $\nu$ and $\nu + d\nu$, assuming it has the Planck spectrum, is given by

$$u_\nu d\nu \equiv \frac{8\pi \nu^2 h\nu\, d\nu}{c^3 [\exp(h\nu/k_B T)-1]}. \tag{20}$$

Therefore, the number density of photons

$$n_\nu = u_\nu d\nu/h\nu = \frac{8\pi \nu^2 d\nu}{c^3 [\exp(h\nu/k_B T)-1]}. \tag{21}$$

The number of photons in a volume $V$ is conserved, i.e.

$n_v V = n_{v_0} V_0 \Rightarrow n_v V_0 (1+z)^{-3} f(z_c)^3 (1+z_t)^3 = n_{v_0} V_0$, or

$$n_v = n_{v_0}(1+z)^3 f(z_c)^{-3}(1+z_t)^{-3}. \tag{22}$$

We may now write,

$$n_v = \frac{8\pi v_0^2 \, dv_0}{c_0^3 \, [\exp(h_0 v_0 / k_{B,0} T_0) - 1]} (1+z)^3 f^{-3} (1+z_t)^{-3},$$

$$= \frac{8\pi v^2 \, dv}{c^3 \, [\exp(hv / k_B T) - 1]} (1+z_t)^{-3}, \tag{23}$$

where $T = T_0(1+z)$. We can see that the CMB emission has the black body spectrum, but its intensity scales as $(1+z_t)^{-3}$ due to the tired-light effect. We can now determine $z_{ls}$ in the next section.

## III. RESULTS

Let us first see how the critical density in the CCC+TL model is related to that in the standard ΛCDM model. Using Eq. (6), $\varepsilon_{c,0}^{CCC+TL} / \varepsilon_{c,0}^{\Lambda CDM} = (H_c + \alpha)^2 / H_0^2$. From Gupta (2023), by fitting the supernovae type 1a data (Pantheon+), $H_c = 59.51$, $\alpha = (-0.7997)H_c$, and $H_0^{Planck} = 67.4$. Thus, $\varepsilon_{c,0}^{CCC+TL} = 0.031 \varepsilon_{c,0}^{\Lambda CDM}$, which is about the same as the estimated baryon density from observation of the visible matter (Ryden 2017). Next, we need to determine the current radiation energy density in the CCC+TL model. It is estimated as $\Omega_{r,0}^{\Lambda CDM} \sim 9 \times 10^{-5}$ in the ΛCDM model (Ryden 2017), including neutrinos. Therefore, in the CCC+TL model, it is $\Omega_{r,0}^{CCC+TL} \sim 2.9 \times 10^{-3}$, and $\Omega_{b,0}^{CCC+TL} = 1$.

*Sound Horizon*: We know that the CMB temperature $T_0 = 2.7255$ K. For $T \approx 3000\, K$ required for the surface of the last scattering, we take $1 + z = (1 + z_c)(1 + z_t) = 1091$. Now, only the expanding universe is responsible for the sound horizon and CMB power spectrum, i.e., we need to know the value of $z_c$ when $z = 1090$, which is then taken as $z_{ls}$. Following the steps in (Gupta 2023), we determined $z_{ls} = 166$. Briefly, Eq. (44) in the paper relates $z_c$ and $z$. It is solved numerically for $z_c$ for any value of $z$, given $H_c$ and $\alpha$, keeping in mind that the function $f$ is also expressed in terms of $z_c$, $H_c$ and $\alpha$. We used 'fzero' function in Matlab for the purpose. Once we have $z_c$, $z_t$ is obtained using $(1+z) = (1+z_c)(1+z_t)$.

The angular diameter distance $d_A$ in the CCC+TL model (Gupta 2023)

$$d_A(z_c) = f(z_c) d_p(z_c) / (1 + z_c). \tag{24}$$

The angular size of the sound horizon is $\theta_{sh} = d_{sh}(z_{ls})/D_A$. Using Eqs. (15) to (18), the computed values, with $z_{ls} = 166$ and baryon density $\Omega_{b,0} = 1$, are $\theta_{sh} = 0.60°$ as measured by Planck (Planck Collaboration 2020), $d_{sh} = 15.5$ Mpc, and $D_A = 1.49$ Gpc, when we set $\alpha = -0.845 H_x$. The uncertainties in these values are not estimated as they are not relevant to our analysis of the model.

*BAO Acoustic Scale*: We will now examine BAO measurement from the angular separation of pairs of galaxies in thin redshift bins. Carvalho et al. (2021) present the data in $z \in [0.105, 0.115]$ identified as $z = 0.11$ bin. Lemos et al. (2023) provided them in 11 thin redshift bins ranging from $z = 0.44$ to $z = 0.66$ based on the work of Carvalho et al. (2016) and Carvalho et al. (2020). We may estimate the angular BAO scale $\theta_{BAO}(z)$, when using thin redshift bins, with the expression (Lemos et al. 2023),

$$\theta_{BAO}(z) = \frac{r_s}{(1+z) d_A(z)} = \frac{r_s}{d_p(z)} \Rightarrow r_s = d_p(z) \theta_{BAO}(z), \tag{25}$$

where $r_s$ is the BAO characteristic length scale and $d_A(z) = d_p/(1+z)$ for the standard model. We therefore used Eq. (25) with $d_p$ calculated for each redshift bin to determine $r_s$ and compared it with its expected value of $r_s \approx 150$ Mpc. We therefore used Eq. (25) with $d_p$ calculated for each redshift bin to determine $r_s$ to see how it differs from its expected value of $r_s \approx 150$ Mpc. We present the results in Table 1 for the CCC+TL model. The uncertainties in the $r_s$ values correspond to the uncertainties in the $\theta_{BAO}$ values. One could expect the uncertainties to reduce similarly to those reported by Lemos et al. (2023) if the Gaussian Process method was used to reconstruct the binned supernovae data. The values in Table 1 were fitted with straight lines using Matlab's 'Curve Fitting' tool and presented in Figure 1. The weighted average value of $r_s = 151.0$ ($\pm 5.1$) Mpc was obtained by fitting the same values by constraining the fit line with zero slope; the 95% confidence bounds are shown in parentheses. When we did the same calculations for the ΛCDM model with the parameters $H_0 = 72.99$ Km s$^{-1}$ Mpc$^{-1}$ and $\Omega_{m,0} = 0.351$ (Gupta 2023), we got the results shown in Figure 2, and $r_s = 145.2$ ($\pm 5.0$) Mpc. Considering the error bars, the slopes in the figures may not be meaningful. The values for the two models agree within their 95% confidence bounds.

TABLE 1. BAO data from the angular separation of pairs of galaxies fitted to extract the absolute scale of BAO.

| $z_{bin}$ | $\Delta z_{bin}$ | $\theta_{BAO}(z_{bin})°$ | $z_c$ | $d_p$ (Mpc) | $r_s$ (Mpc) |
|---|---|---|---|---|---|
| 0.11 | [0.105, 0.115] | 19.8±1.05 | 0.088 | 439 | 151.7±8.0 |
| 0.45 | [0.44, 0.46] | 4.77±0.17 | 0.347 | 1684 | 140.2±5.0 |
| 0.47 | [0.46, 0.48] | 5.02±0.25 | 0.362 | 1752 | 153.5±7.7 |
| 0.49 | [0.48, 0.50] | 4.99±0.21 | 0.376 | 1819 | 158.4±6.7 |
| 0.51 | [0.50, 0.52] | 4.81±0.17 | 0.391 | 1886 | 158.3±5.6 |
| 0.53 | [0.52, 0.54] | 4.29±0.30 | 0.405 | 1952 | 146.1±10.2 |
| 0.55 | [0.54, 0.56] | 4.25±0.25 | 0.419 | 2018 | 149.7±8.8 |
| 0.57 | [0.56, 0.58] | 4.59±0.36 | 0.433 | 2083 | 166.9±13.1 |
| 0.59 | [0.58, 0.60] | 4.39±0.33 | 0.448 | 2148 | 164.6±12.4 |
| 0.61 | [0.60, 0.62] | 3.85±0.31 | 0.462 | 2211 | 148.6±12 |
| 0.63 | [0.62, 0.64] | 3.90±0.43 | 0.476 | 2276 | 154.9±17.1 |
| 0.65 | [0.64, 0.66] | 3.55±0.16 | 0.490 | 2339 | 144.9±6.5 |

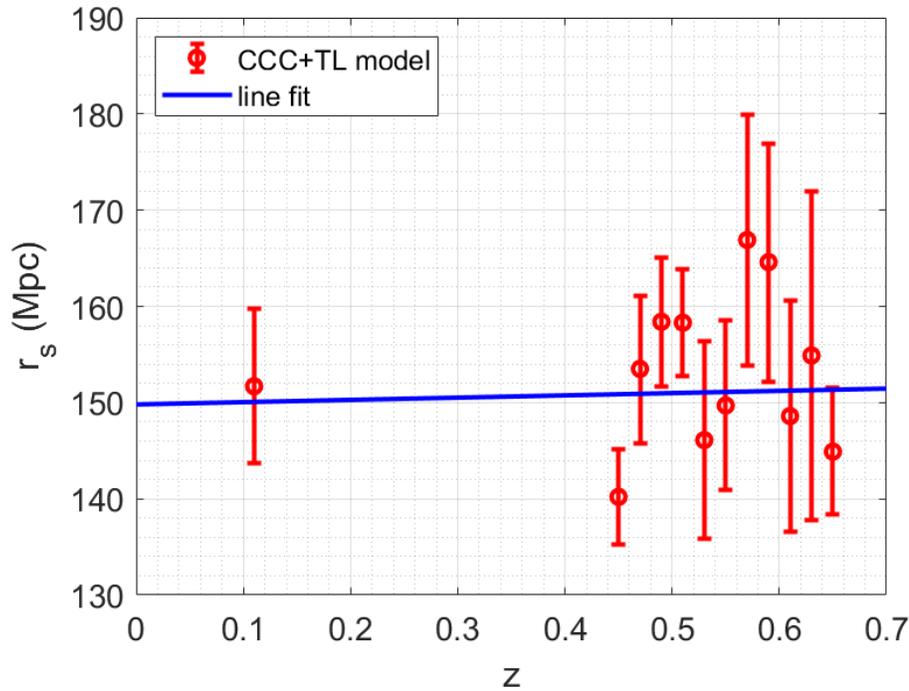

Figure 1. The absolute scale of BAO estimated at different redshifts using the CCC+TL model.

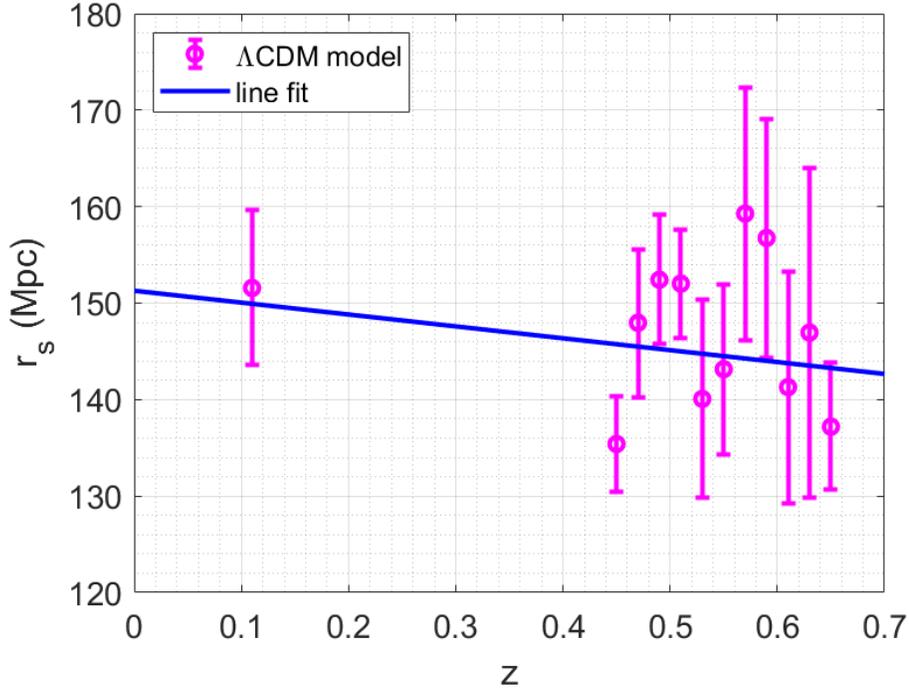

Figure 2. The absolute scale of BAO estimated at different redshifts using the ΛCDM model.

One would also like to know the sensitivity of the $r_s$ with respect to the changes in $H_c$ and $\alpha$. With 95% confidence (2σ), we have $H_c = 59.51(\pm 1.06)$ Km s$^{-1}$ Mpc$^{-1}$ and $\alpha = -0.7997(\pm 0.0143)H_c$ (Gupta 2023). For $z_{bin} = 0.11$, the value of $r_s$ increases from 149.1 to 154.5 Mpc with decreasing $H_c$ over 2σ for $\alpha = -0.7997H_c$, and $r_s$ varies only 0.2 Mpc with varying $\alpha$ over 2σ for all the $H_c$ values. Correspondingly, for $z_{bin} = 0.65$, the value of $r_s$ increases from 142.5 to 147.6 Mpc with decreasing $H_c$, and varies 1.2 Mpc with varying $\alpha$ over 2σ. It means that the $r_s$ values are relatively stable against the changes in $H_c$ and $\alpha$, especially when compared to the uncertainties in their value due to uncertainties in $\theta_{BAO}$ values.

As a side note, it can be shown that the Jean density in the new model decreases by a factor of $f^3$ (~30 $at$ $z = 20$) as compared to the standards model and, therefore, star formation can begin at higher temperature and redshift by a factor of $f$. Also, the matter-radiation equality happens at $z \sim 10,000$.

## IV. DISCUSSION

The primary objective of this paper is to see if the recently proposed cosmological model CCC+TL (Gupta 2023) is compliant with (a) the sound horizon angular size $\theta_{sh}$ resulting from the oscillations in the primordial baryon-photon plasma that is imprinted and observed in the CMB thermal anisotropy spectrum; and (b) the absolute scale $r_s$, believed to be resulting from the baryon acoustic oscillations, as derived from the measurements of the angular size of the two-point correlation function $\xi(s)$ of the separation $s$ of millions of galaxies. Most models assume these are correlated. However, in the words of Sutherland (2012): "The baryon acoustic oscillations (BAO) feature in the distribution of galaxies provide a fundamental standard ruler which is widely used to constrain cosmological parameters. In most analyses, the comoving length of the ruler is inferred from a combination of CMB observations and theory. However, this inferred length may be biased by non-standard effects in early universe physics; this can lead to biased inferences of cosmological parameters …". Since our model is different from the standard models, we will not attempt to correlate the two.

Currently, we do not have any CCC+TL parameters by fitting the CMB thermal anisotropy spectrum. Thus, we decided to test the new model for its consistency with the CMB sound horizon and the BAO low redshift observations using the model parameters determined by fitting the Pantheon+ data (Scolnic et al. 2022; Brout et al.

2022): $H_c = 59.51\ (\pm 1.06)$ Km s$^{-1}$Mpc$^{-1}$ and $\alpha = -0.80\ (\pm 0.01)\ H_c$. These are the only two parameters required for fitting the data. Ignoring the uncertainty, they result in $H_0 = 72.62$ Km s$^{-1}$Mpc$^{-1}$ (Gupta 2023).

The only adjustable parameter when fitting the sound horizon angular size at recombination (Eqs. 15 to 18) has been taken to be the covarying coupling constant parameter $\alpha$. It is because the function chosen to define covarying coupling constants is arbitrary and should not be expected to remain unchanged over the full cosmological timescale. Fortunately, only 5.6% change in $\alpha$ is required to yields the angular size $\theta_{sh} = 0.60°$ from Planck observations (Planck Collaboration 2020 – Table 6). It is only about half the Hubble tension. However, the model-dependent physical size $d_{sh} = 15.5$ Mpc is very different from the ΛCDM model value of $d_{sh} \cong 135$ Kpc using Planck parameters: $H_0 = 67.32$ Km s$^{-1}$Mpc$^{-1}$, $\Omega_{m,0} = 0.3158$, and $\Omega_{b,0} = 0.16\Omega_{m,0}$; the ratio of the two is $\sim 100$. The main reason for such a significant difference in $d_{sh}$ values is that the expanding universe redshifts of the last scattering surface for the two models are vastly different: the expanding universe component for the CCC+TL model is 166 versus 1089 for the ΛCDM model, i.e., the sound horizon progresses for a longer duration in the CCC+TL compared to the ΛCDM model. Since the angular diameter distance $D_A$ at the last scattering surface for the CCC+TL model is ~100 times larger than for the ΛCDM model, about the same as the ratio of the $d_{sh}$ for the two models (Figure 3), $\theta_{sh} (= 0.60°)$ is the same for the two models. The figure is drawn with $\alpha = -0.845 H_x$.

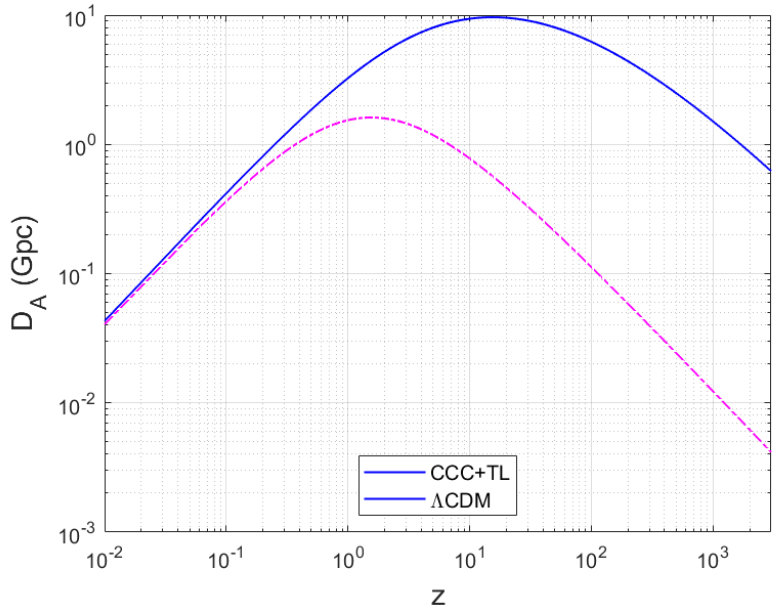

Figure 3. Angular diameter distance as a function of redshift in CCC+TL and ΛCDM models.

The baryon density in the new model is the same as the critical density. However, since there is no *explicit* dark energy in the new model, all of the critical density at present is comprised of matter in the CCC models. Since the critical density in the CCC+TL model is only 3.1% of the standard model, it can all be accounted by the baryonic matter in the universe.

We have shown that the CMB emission has the black body spectrum in the CCC+TL model. Its intensity scales as $(1 + z_t)^{-3}$ due to the tired-light effect. Since $z_t = 0$ at present, we see the evolving spectrum as a perfect blackbody when observed.

The question naturally arises if an experiment can be designed to test the existence of the two redshift components. Wang et al. (2022, 2023) have proposed a method of testing redshift drift, a direct measure of cosmological expansion, using gravitationally lensed images of distant objects. Multiple images arriving at the same instant would traverse different distances and thus would have left an object at different times. With high enough accuracy in measuring redshifts, one could estimate the redshift drift using any two images, hence the cosmological expansion. Any difference in the values of expansion rate calculated using different pairs of object images could be due to the tired light redshift. Another gravitational lensing method uses the supernovae type 1a with a well-defined brightness-peaking profile or quasars with some brightness fluctuation profile marker (Cao et al. 2018, 2020; Gupta 2021a). Such markers in their luminosity show up at different times in their images; different path lengths would mean different contributions to the redshift due to the tired light effect, while the redshift due to the expansion would be the same in such images.

One could also consider designing an experiment to test the coupling constants' variation. One has to be cognisant that in the CCC model, variation of several coupling constants is determined by a single function, i.e., fixing one coupling constant leads to this function not varying and thus all coupling constants not varying. We have yet to find an observation or experiment that can detect the variation of this function and, therefore, of the coupling constants (Gupta 2021a, 2021b, 2022b).

It is perhaps worth mentioning that the globular cluster age determination may not be able to validate or falsify a model based on its prediction of the universe's age. As is well known, the age of globular clusters is model-dependent, and models have been adjusted whenever the age of a star or cluster exceeds the universe's age. Bolte and Hogan (1995) determined certain cluster ages $15.8 \pm 2.1$ Gyr. Considering the Methuselah star, Tang and Joyce (2021) revised its best age estimate from $14.5 \pm 0.8$ Gyr down to a comfortable $12.0 \pm 0.5$ Gyr by adjusting parameters in the MESA stellar evolution code. If the universe's age is established by other methods to be significantly higher than the currently accepted 13.8 Gyr, astrophysicists will be relieved from the burden of constraining stellar ages below 13.8 Gyr. This is evident from the work of Llorente de Andrés (2023), who determined the age of globular cluster NGC104 between 19.04 and 20.30 Gyr after learning that the universe could be 26.7 Gyr old. With no age constraint to worry for recently born clusters, Jeffries et al. (2023) adjusted the age of a young open cluster IC 4665 from 32 Myr to greater than 50 Myr. Thus, the age of a star or a cluster cannot be considered a constraint on the universe's age.

Admittedly, the CCC+TL model is significantly more complex to work with than the ΛCDM model. However, extending simple models to account for precision observations leads to tensions. In the age of precision cosmology, we need to be vigilant about new models that may be needed to go beyond the domain of cosmology so eloquently serviced by the standard model. A word of caution: applied hastily, partially, or incorrectly, CCC+TL would lead to wrong results; it needs good account keeping of all that might be affected when moving from the standard model to the CCC+TL model.

## V. CONCLUSION

The CCC+TL model has successfully resolved the 'impossible early galaxy' problem by stretching rather than compressing the timeline for the formation of stars and galaxies as required by the ΛCDM model. The resulting almost doubling in the age of the universe and increasing the formation times by an order of magnitude has been a subject of concern and requires that the new model also explain some critical cosmological and astrophysical observations, such as CMB, BBN elemental abundances, and BAO. We have presented in this paper, using the new model, the calculation of i) the low-redshift BAO absolute scale, which is the same as observed and estimated using the standard model within the 95% confidence level of the two models, and ii) the sound horizon angular size consistent with Planck observation at the surface of last scattering and established that all the critical density comprises the baryon density with no room for dark matter. Due to the involvement of covarying coupling constants and the hybrid nature of the CCC+TL model, the BAO feature in the tracer power spectrum (CMB) is not at the same scale as in the matter (galaxies) power spectrum (Dodelson & Schmidt 2021). We now have additional confidence to continue with the development of CMB and BBN codes tailored to the new model for testing it further.

**Acknowledgment**

The author is grateful to the reviewer of the paper for constructive comments which improved the quality and clarity of the paper, and to Professors Brian Keating, Avi Loeb, Nikita Lovyagin, and David Spergel for communications with him and constructive critical comments on the CCC+TL model. He wishes to thank Macronix Research Corporation for the unconditional research grant for this work.

**Data availability**

References have been provided for the data used in this work.